\documentclass[conference]{IEEEtran}
\IEEEoverridecommandlockouts
\usepackage{cite}
\usepackage [english]{babel}
\usepackage [autostyle, english = american]{csquotes}
\MakeOuterQuote{"}
\usepackage{amsmath,amssymb,amsfonts}
\usepackage{pifont}
\newcommand{\cmark}{\ding{51}}%
\newcommand{\xmark}{\ding{55}}%
\usepackage{amsthm}
\newtheorem{definition}{Definition}
\usepackage[linesnumbered,ruled,vlined]{algorithm2e}
\usepackage{graphicx}
\usepackage{textcomp}
\usepackage{dblfloatfix}
\usepackage{tablefootnote}
\usepackage{url}
\usepackage{booktabs}
\usepackage{multirow}
\usepackage{comment}
\usepackage{makecell}
\usepackage{xcolor}
\usepackage{tablefootnote}
\usepackage{float}
\usepackage{pgfplots}
\usepackage{tabularray}
\usepackage{balance}
\usepackage[usestackEOL]{stackengine}
\pgfplotsset{width=7cm,compat=1.9}
\def\BibTeX{{\rm B\kern-.05em{\sc i\kern-.025em b}\kern-.08em
    T\kern-.1667em\lower.7ex\hbox{E}\kern-.125emX}}

\begin{document}

\title{TELSAFE: Security Gap Quantitative Risk Assessment Framework \\
}

\author{\IEEEauthorblockN{Sarah Ali Siddiqui\IEEEauthorrefmark{1}\thanks{E-mail:sarah.siddiqui@data61.csiro.au}, Chandra Thapa\IEEEauthorrefmark{1}, Derui Wang\IEEEauthorrefmark{1}, Rayne Holland\IEEEauthorrefmark{1}, Wei Shao\IEEEauthorrefmark{1}, Seyit Camtepe\IEEEauthorrefmark{1}, Hajime Suzuki\IEEEauthorrefmark{1}\\ and Rajiv Shah\IEEEauthorrefmark{2}}\\
\IEEEauthorblockA{\IEEEauthorrefmark{1}CSIRO Data61, Sydney, Australia}
\IEEEauthorblockA{\IEEEauthorrefmark{2}MDR Security, Canberra, Australia}
}

\maketitle


\begin{abstract}
Gaps between established security standards and their practical implementation have the potential to introduce vulnerabilities
, possibly exposing them to security risks. To effectively address and mitigate these security and compliance challenges, security risk management strategies are essential. 
However, it must adhere to well-established strategies and industry standards to ensure consistency, reliability, and compatibility both within and across organizations. 
%
In this paper, we introduce a new hybrid risk assessment framework called TELSAFE, which employs probabilistic modeling for quantitative risk assessment and eliminates the influence of expert opinion bias. 
%
 The framework encompasses both qualitative and quantitative assessment phases, facilitating effective risk management strategies tailored to the unique requirements of 
organizations. A specific use case utilizing Common Vulnerabilities and Exposures (CVE)-related data demonstrates the framework's applicability and implementation in real-world scenarios, such as in the telecommunications industry.
%
%

\end{abstract}


\section{Introduction}
\label{sec:intro}

With the rise in security breaches, companies must continuously evaluate their security posture, identify weaknesses, and implement remediation strategies to minimize potential threats in a structured and prioritized manner. By conducting thorough risk\footnote{\emph{Note: The terms risk, risk management and risk assessment refer to security-specific risk, security-specific risk management, and security risk assessment, respectively, throughout this paper.}} management, businesses can gain insights into the vulnerabilities present in their IT environment and develop strategic plans to mitigate these risks effectively.
As organizations increasingly depend on IT systems, applications, and digital assets, the significance of proactive risk assessments continues to grow, ensuring resilience against evolving cyber threats \cite{intro29}. The key components of the risk management process include context establishment, risk assessment, risk mitigation, and risk monitoring \cite{intro18}. Risk assessment is one of the major steps in the risk management process. Risk assessment guides decisions in scenarios like prioritizing threat mitigation and/or resource allocation for addressing security issues\cite{intro21, intro22}. Risk can be assessed (i) qualitatively, where subjective opinions are analyzed to evaluate risk \cite{intro23}, or (ii) quantitatively, where numerical data and statistical methods are employed to quantify risks \cite{intro24}, or (iii) with a hybrid approach, by combining both qualitative and quantitative techniques to measure risk \cite{intro25, intro26, intro27}. Qualitative analysis is important to incorporate the vision of a specific organization in terms of its requirements, goals, and the nature of the business. On the other hand, quantitative evaluation helps make decisions based on objective criteria rather than subjective opinions. Moreover, it is important to quantify the risk to estimate the severity of the risk and eliminate subjectivity and inconsistencies in evaluating these risk levels to make informed decisions \cite{intro28}. It is crucial to comply with industry standards while developing risk management strategies, as industry standard-compliant security solutions ensure consistency, reliability, and compatibility within and across organizations. The industry standards that guide risk management, in general, for information security 
include ISO 31000, ISO 31010, and ISO 27005
\cite{intro17, intro18, intro19}
. An effective and standards-compliant risk assessment module consists of three major components, including risk identification, risk analysis, and risk evaluation \cite{intro17}. 

\begin{figure}[b]
\centering
    \includegraphics[width=0.75\linewidth]{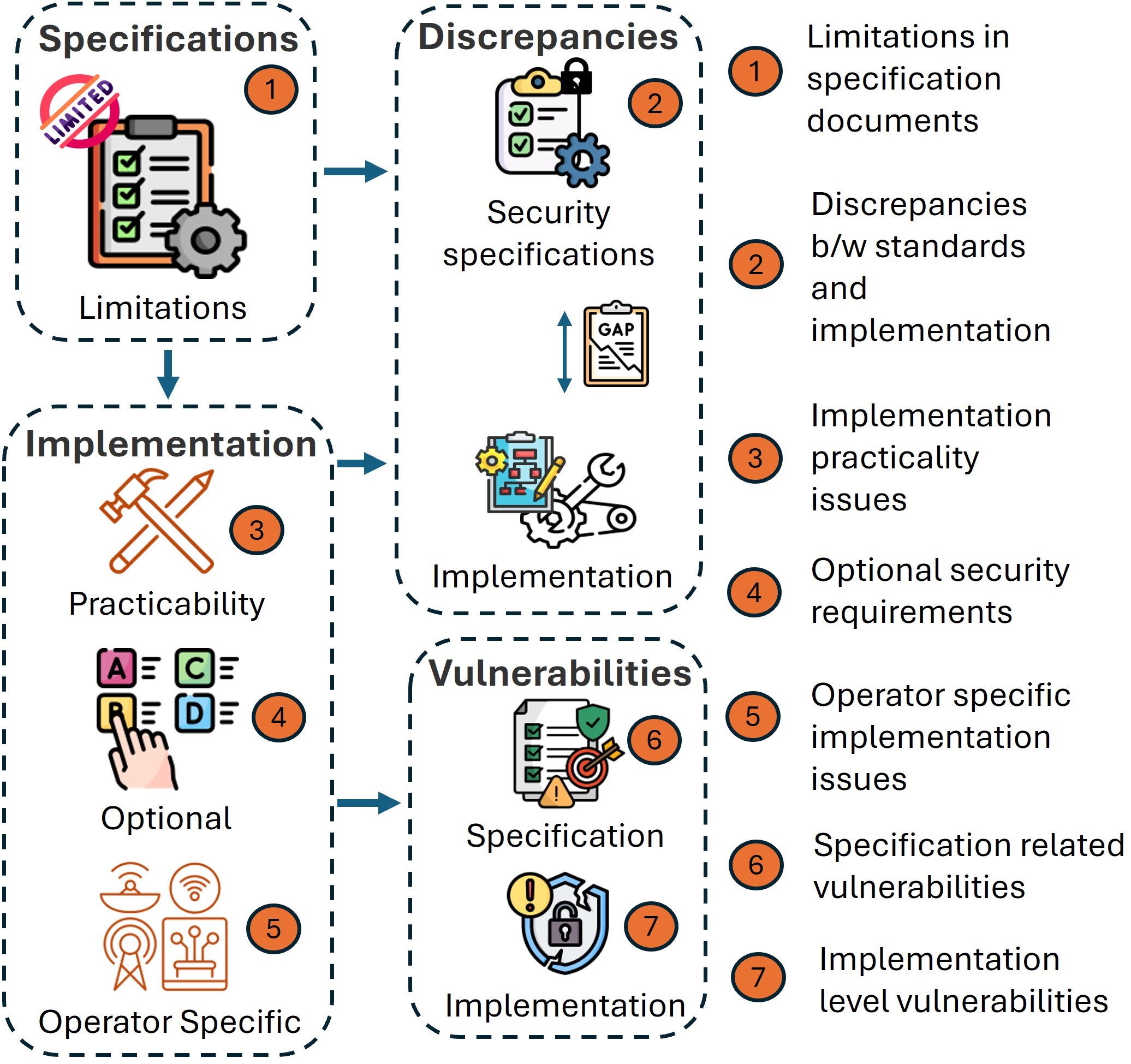}
    \caption{Potential Gaps -- Standards and Implementation.}
    \label{fig:potential_gaps}
\end{figure}

\begin{figure*}[b]
\centering
    \includegraphics[width=0.8\linewidth]{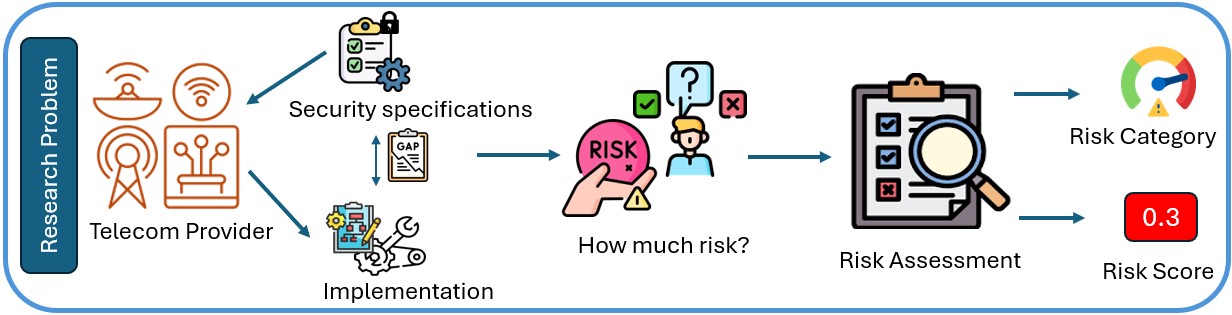}
    \caption{Research Problem: Risk Assessment for Gaps between Standards' Security Specifications and Implementation, including Use case Scenario for Telecommunication Industry.}
    \label{fig:research_problem}
\end{figure*}


While developing or adopting specific risk assessment methodologies, it is important that the designed approaches employ specific techniques for each step appropriate for an organization instead of overly generic guidelines, as guidelines can be found in standards' documents \cite{intro17}. 
Concurrently, these frameworks need to be generic enough to cover various information security risks. 
Risk assessment can be applied across various applications and specific focus areas. The results of individual risk assessments conducted in different areas of interest can be integrated to provide a comprehensive risk evaluation of the entire system or organization.

\begin{definition} [\textbf{Risk}]
    Risk is commonly characterized by the sources of risk, potential events, their consequences, and the likelihood of their occurrence, wherein likelihood refers to the possibility of an event happening, whereas consequences refer to the impact if such an event were to occur \cite{intro19}. 
\end{definition}

Several standardization organizations, e.g., International Organization for Standardization (ISO), 3rd Generation Partnership Project (3GPP), and 
International Electrotechnical Commission (IEC), are actively engaged in improving the security of 
networks \cite{intro1, intro2}. 
One of their challenges is gaps in standards' specifications and vendor/operator implementation. These gaps exist due to (i) limitations in the standard specifications and/or implementations \cite{intro2, intro4}, (ii) the interpretation of vague security requirements \cite{intro5}, (iii) discrepancies between standards and their implementation \cite{intro6}, and (iv) implementation level vulnerabilities \cite{intro7, intro8}. The gap may also exist due to practicality issues. For example, although new security specifications have been defined to enhance the security of a system
, not all security improvements outlined in these specifications can fulfill their potential in practical application. Moreover, some security requirements are optional instead of obligatory, allowing vendors/operators to overlook them. Even when security requirements are obligatory, stakeholders might choose to bypass them for practical considerations like performance, cost, and compatibility~\cite{intro10}. 
Figure \ref{fig:potential_gaps} identifies potential gaps related to standards and implementation in a 
system.
The gaps can lead to vulnerabilities and create security risks, e.g., unauthorized access, information leakage, and service unavailability \cite{intro6, intro11, intro12, intro13, intro14, intro15, intro16}. 
Effective security risk management is essential to mitigate these vulnerabilities and ensure compliance with industry standards.

%

\begin{table*}[b]
\centering
\caption{Comparison with Existing Literature}%
\label{tab:comparison_sota}
{\renewcommand{\arraystretch}{0.5}
\scriptsize{
\begin{tabular}{p{1.6cm}p{1.2cm}
p{2.2cm}p{1.75cm}p{2.7cm}p{6cm}}
\toprule
\textbf{Framework} & \textbf{Quantitative} 
& \textbf{Risk Type} & \textbf{Technique} & \textbf{Dataset} & \textbf{Limitations \& Challenges}\\
\tabularnewline
\midrule
\centering{CyRiPred \cite{intro31}} & \centering{\cmark} 
& Cyber risk categories & LLMs & CVE-based & Wikipedia, attack surface, only the broader categories have been validated instead of each CVE  \\\tabularnewline
\centering{\cite{intro29}}  & \centering{\cmark} 
& Cyber security & CVSS metrics & \xmark & CVSS metrics only, no validation has been performed  \\\tabularnewline
\centering{\cite{sota39}} & \centering{\cmark} 
& Information security & Probabilistic & OWASP top 10 & Only 10 sample dataset, lacks a clear explanation of how the values for likelihood and impact are estimated, no validation has been performed 
\\\tabularnewline
\centering{\cite{intro30}}  & \centering{Semi} 
& Privacy & Fuzzy logic & \xmark & Reliance on expert opinions, subjective membership functions, pre-defined weights, no validation has been performed \\\tabularnewline
\centering{AFPr-AM \cite{sota40}} & \centering{Semi} 
& Privacy & Fuzzy AHP and game theory & Via questionnaires & Reliance on expert opinions for ranking factors, pre-defined weights, no validation has been performed \\\tabularnewline
\centering{\cite{intro32}} & \centering{\xmark} 
& Cyber security & AI-based &  Via questionnaires & Focus on AI impact, attack surface, ethical concerns, no validation has been performed \\\tabularnewline
\centering{TELSAFE
} & \centering{\cmark} 
& Gap in security & Probabilistic &  CVE-based (implementation illustration only) & No validation dataset is publicly available 
\\\tabularnewline
\bottomrule
\end{tabular}
}}
\end{table*}


A majority of risk assessment models focusing on quantitative assessment employ semi-quantitative methods, such as fuzzy logic, that rely on expert opinions for membership function and weight determination, making them subjective \cite{intro30}. Some use unreliable data sources \cite{intro31}, whereas others focus on AI-based techniques that increase the attack surface and cause concerns regarding non-compliance with GDPR, among other limitations \cite{intro32}. Please note that the argument here is not against embracing AI technologies 
but to limit their application to where it is truly required
to reduce the attack surface. Specific details regarding each work have been presented in Section \ref{sec:literaturereview}. Regarding industry tools, NIST RMF (Risk Management Framework) presents a high-level guideline instead of a specific risk assessment method \cite{intro33}, FAIR (Factor Analysis of Information Risk) focuses on quantitative assessment; however, it only considers security from a financial standpoint \cite{intro34}, COBIT (Control Objectives for Information and related Technology) offers guidelines for managing and optimizing enterprise IT processes instead of a specific risk assessment method \cite{intro35}, and OCTAVE (Operationally Critical Threat, Asset and Vulnerability Evaluation) emphasizes on qualitative risk assessment related to operational risks only \cite{intro36}. More details on the limitations of each of these tools have been included in Section \ref{sec:literaturereview}. 




This paper proposes a risk assessment framework called TELSAFE that adheres to the guidelines provided by the ISO, and IEC 
standards, and it further segments the analysis process into distinct and manageable steps, i.e., qualitative analysis, risk scenario development, and risk modeling. In addition, it includes a context definition step within the risk assessment process to specify the focus area of a particular assessment, which is especially helpful to add flexibility in breaking down complex systems into smaller, more manageable risk assessment areas. Moreover, this framework relies on a hybrid approach that leverages the strengths of both qualitative and quantitative assessment approaches while mitigating their respective limitations through the benefits of the other. It utilizes qualitative assessment for context definition, risk factor identification, and risk analysis to capture an organization's vision, defined by its requirements, objectives, and business nature. For subsequent steps, including risk scenario development, risk modeling, and risk evaluation, it employs a quantitative assessment to eliminate subjectivity by quantifying risk scores associated with the gaps in the standards and implementation. 
Incorporating specific techniques outlined in the standards for particular steps diminishes vagueness and complexity in the adoption of the framework. Due to this simplified adoption and the hybrid approach, it offers a flexible approach that can be adapted to align with the unique requirements, objectives, and risk landscapes/profiles of different organizations. Moreover, a use case 
tailored specifically for CVE (Common Vulnerabilities and Exposures) related information has also been presented to demonstrate the functionality of the envisaged security risk assessment framework on real-world scenarios such as in 5G/6G networks or telecommunications sector, through a CVE dataset. \textit{To the best of our knowledge, standards' specifications and implementation gap-related security risk assessment have not been explored previously}. This work serves as an initial step toward effective risk assessment in the telecommunications industry. The research at hand employs probabilistic modelling (event trees) for risk scenario development during the quantitative assessment phase; section \ref{sec:literature_review} discusses different methods of computation in detail. Probabilistic modelling techniques have been widely employed for assessing safety and security risk assessment in complex systems, e.g., nuclear power plants \cite{sota44}, aerospace \cite{sota45} and maritime autonomous surface ships \cite{sota46}, critical infrastructures, e.g., oil and gas \cite{sota47}, and enterprise systems \cite{sota43}. In addition, the risk analysis framework by NIST \cite{sota43} utilizes probabilistic methods, indicating their effectiveness in enterprise networks. Event trees provide a structured approach for the quantification of likelihood, capturing chain events, and prioritizing risks based on probabilities \cite{addcontri1, introadd3}. 

Given the current threat landscape, achieving efficient and automated designs is nearly impossible without leveraging AI-based techniques such as ML or Large Language Models (LLMs), making it both practical and wise to embrace these advancements. However, it is important to consider the repercussions of the same and keep their involvement to a minimum. Therefore, while organizations may need to employ AI-based techniques for certain aspects of risk assessment, the quantitative assessment proposed in this manuscript does not rely on any AI-driven approaches. However, it is flexible and can be easily extended to incorporate AI methodologies if required. 
Probabilistic models are employed in some industries as they align with the regulatory requirements related to these industries and risk management \cite{introadd2}. Figure \ref{fig:research_problem} illustrates the research problem at hand.

\begin{table*}[b]
\centering
\caption{Comparison with Existing Tools}%
\label{tab:comparison_sota_tools}
{\renewcommand{\arraystretch}{0.5}
\scriptsize{
\begin{tabular}{p{1.5cm}p{1.2cm}
p{2.0cm}p{1.5cm}p{1.7cm}p{7cm}}
\toprule
\textbf{Framework} & \centering{\textbf{Quantitative}} 
& \textbf{Risk Type} & \centering{\textbf{Focused}} & \centering{\textbf{Standard Alignment - ISO 31000}} & \textbf{Limitations}\\ 
\tabularnewline
\midrule
RMF \cite{intro33} & \centering{\xmark} 
& Privacy and information security & \centering{\xmark} & \centering{\xmark} & Extreme complexity in implementation for smaller scale, difficulty in emerging threats, ensuring consistency, slow adaptation, resource limitations addressing, etc. \\\tabularnewline 
FAIR \cite{intro34} & \centering{\cmark} 
& Informational and operational & \centering{Financial} & \centering{\cmark} & Time consuming and complex implementation, data quality and timeliness problem, scalability challenges, etc. \\\tabularnewline
COBIT \cite{intro35} & \centering{\xmark} 
& Management and governance & \centering{\xmark} & \centering{\cmark} & Resource availability for small businesses, implementation complexity, and customization, etc. \\\tabularnewline
OCTAVE \cite{intro36} & \centering{\xmark} 
& Operational & \centering{\xmark} & \centering{\cmark} & Complex and resource-intensive integration process, gaps in addressing all emerging threats, documentation requirements, lengthy implementation, etc. \\\tabularnewline
EU \cite{intro37} & \centering{\xmark} 
& -- & \centering{\cmark} & \centering{\xmark} & Current implementation status is unclear \\\tabularnewline
ANU \cite{intro38} & \centering{\xmark} 
& -- & \centering{\cmark} & \centering{\cmark} & Project in progress \\\tabularnewline
Proposed & \centering{\cmark} 
& Gap in security & \centering{\cmark} & \centering{\cmark} & -- \\\tabularnewline
\bottomrule
\end{tabular}
}}
\end{table*}




\subsection{Our Contributions}
The research at hand focuses on the following main contributions:

\begin{itemize}

    \item  \textbf{TELSAFE}: Envisaging a hybrid security risk assessment framework, \textit{TELSAFE} which employs specific techniques based on \cite{addcontri1, addcontri2}, especially for quantitative assessment, which reduces the complexity of implementation as opposed to the existing generic frameworks such as NIST RMF \cite{intro33}. 

    \item  \textbf{Adaptability}: Offering better adaptability for unique requirements, objectives, and risk profiles of any organization by utilizing a hybrid approach to leverage numerical assessment while considering qualitative factors as well. 
    
     \item \textbf{Gap related Risk Assessment}: Including risk assessment for a specific application for standards' specifications and implementation gap, which has not been undertaken previously to the best of our knowledge. These gaps may lead to chain events/failures, making a structured approach essential for assessing the probability of such event sequences.
   
    \item \textbf{Probabilistic Modeling}: Employing probabilistic modeling and eliminating the influence of expert opinion bias to enable a quantitative assessment of risks without relying on subjective expert judgments. Probabilistic modeling is widely employed for risk assessment in safety risk management; however, it is underexplored in the security risk assessment domain. Moreover, utilizing event trees for a structured approach to quantify likelihood and capture chain events.  

    \item \textbf{Implementation}: Presenting a specific use case, \textit{CVE-TELSAFE}, to demonstrate the application of the proposed framework on 5G/6G based telecom industry by leveraging a dataset with CVE-related information for quantifying security risks due to the lack of publicly available data on logged security incidents or maintained risk registers.



    
\end{itemize}

\subsection{Organization of the Paper}
The rest of the paper is organized as follows. Section~\ref{sec:literaturereview} reviews the state-of-the-art risk assessment models and tools. Section~\ref{sec:proposedmethod} presents the details of our proposed TELSAFE system architecture and the mathematical model, whereas Section~\ref{sec:use_case} covers a use case scenario for the telecommunications industry based on CVE information. 
Finally, Section~\ref{sec:conclusion} concludes the paper and outlines future directions.



\section{Related Work}
\label{sec:literaturereview}
Security risk management has been discussed in detail by various standards from different standardization organizations; however, almost all of these present generic and high-level guidelines for security risk management without proposing specific methodologies \cite{intro17, intro18, intro20}. This section presents a detailed literature review of the existing risk assessment works in both academia and industry.


\subsection{Literature Review}
\label{sec:literature_review}
In the academic community, quantitative security risk management or assessment is still under-explored.

Kia et al. \cite{intro31} present a CVE-based quantitative risk prediction scheme for cyber risks. The limitations include reliance on Wikipedia for risk category occurrences and likelihood estimation as Wikipedia might not be up to date on certain emerging risks, and it might contain inaccurate or incomplete information, adoption of LLMs may increase the attack surface as they introduce vulnerabilities of their own which is specially challenging due to a lack of risk management related security requirements specific to LLMs, the computation time has not been mentioned, and compliance with related security standards has not been discussed. Aksu et al. \cite{intro29} employ CVSS metrics to compute security risk for IT systems quantitatively; however, the proposed method relies solely on CVSS metrics for risk assessment. 
Padur et al. \cite{sota39} amalgamate both qualitative and quantitative security risk management approaches; however, the dataset consists of 10 samples only, and the proposed approach lacks a clear explanation of how the values for likelihood and impact are estimated. 
Hart et al. \cite{intro30} suggest a risk assessment methodology for privacy risks relying on fuzzy logic. The suggested approach is limited only to privacy risks, employs fuzzy logic, which utilizes subjective membership functions requiring expert opinions, and uses pre-defined weights instead of data-driven weights
. Ahvanooey et al. \cite{sota40} present a privacy risk assessment methodology for social media platforms based on fuzzy AHP (Analytical Hierarchy Process). However, the developed approach relies on expert opinions for ranking factors and determining corresponding weights, which can introduce bias, subjectivity, and inconsistencies. 
Ebere-Uneze et al. \cite{intro32} propose an AI-based cyber risk management approach. However, the authors highlight that the use of AI systems raises ethical concerns and potential risks of non-compliance with the General Data Protection Regulation (GDPR), the adoption of AI has expanded the attack surface, the threat of adversarial attacks exists, where attackers deliberately target, and poison AI datasets and AI may jeopardize confidentiality, and there is a risk of weaponization by malicious actors, e.g., WormGPT. Please note that the authors are not claiming that probabilistic models are immune to attacks. Instead, they are considered more resilient due to their ability to account for uncertainty. By incorporating uncertainty into their predictions, these models can offer a probabilistic evaluation of outcomes, which may make it harder for adversaries to create perturbations that reliably produce erroneous results. 



%

A variety of computational techniques, e.g., fuzzy logic \cite{sota41}, game theory \cite{sota40}, machine learning (ML) \cite{sota42}, large language models (LLMs) \cite{intro31}, and probabilistic methods \cite{sota43}, on their own or in combination with other techniques are employed for security risk assessment \cite{intro19}. Fuzzy logic handles uncertainty and imprecision in risk assessment by using degrees of truth rather than precise values, making it suitable for qualitative scenarios 
but less effective in scenarios requiring objective numerical outcomes. Moreover, the definition of member functions often requires expert opinions, and pre-defined weights are employed instead of data-driven weights, making them subjective \cite{sotaadd1, sotaadd2, sotaadd3}. Game theory is primarily used to model strategic interactions between decision-makers, offering insights into optimal choices in competitive or cooperative environments but lacking straightforward numerical risk scores. AI-based techniques, such as ML and LLMs, expand the attack surface, give rise to ethical concerns and possible non-compliance with GDPR, introduce the threat of adversarial attacks, may threaten confidentiality, and there is a possible concern of exploitation by malicious entities, e.g., WormGPT. Probabilistic risk assessment techniques, where the goal is to acquire a numerical risk score, offer a precise and quantifiable approach to risk management \cite{sotaadd4}. These techniques assess the likelihood of specific risks and their potential impacts to provide a clear numerical representation of risk severity.

Probabilistic techniques are used extensively for safety and/or security risk assessment in complex systems, e.g., nuclear power plants \cite{sota44}, aerospace \cite{sota45} and maritime autonomous surface ships \cite{sota46}, critical infrastructures, e.g., oil and gas \cite{sota47}, and enterprise systems \cite{sota43}. Moreover, NIST's risk analysis framework \cite{sota43} employs probabilistic methods, which suggest that probabilistic methods are effective in enterprise networks. However, security risk management employing probabilistic techniques recommended by the ISO/IEC 31010 standard 
is still under-explored.


\subsection{Existing Industry Tools}
\label{sec:existing tools}
\begin{figure*}[t]
\centering
    \includegraphics[width=0.80\linewidth]{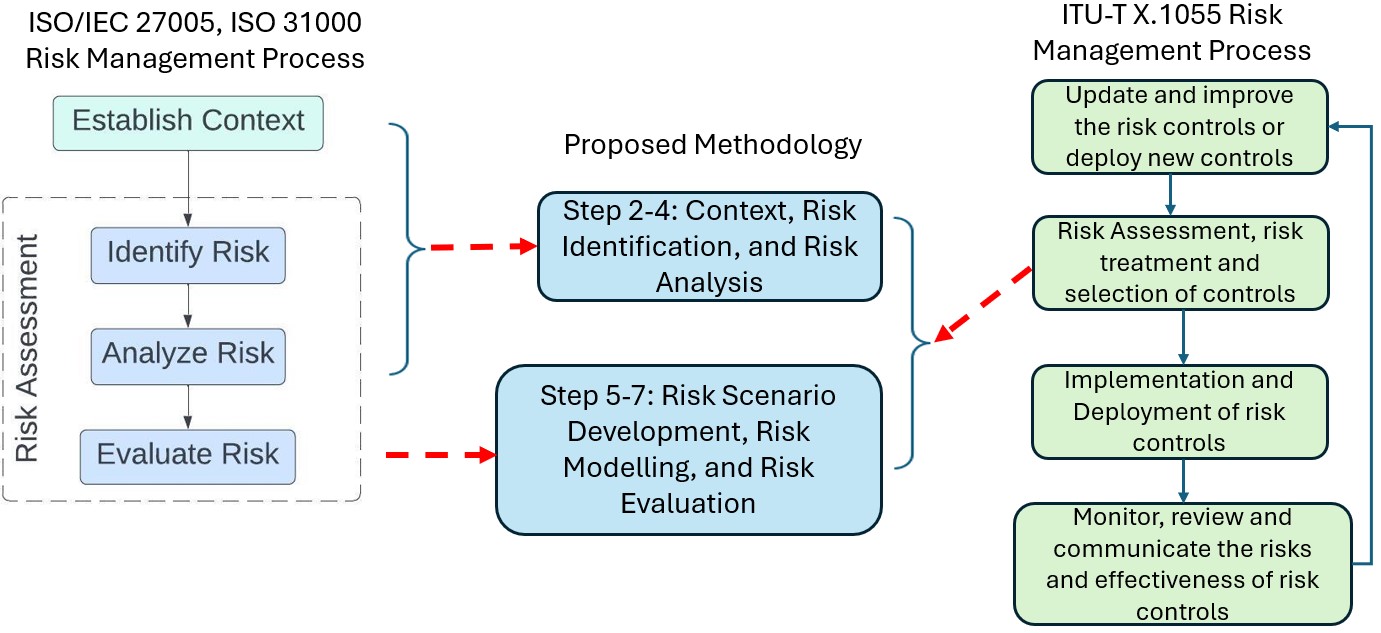}
    \caption{Alignment of the Proposed Risk Assessment Framework with the Existing Risk Management Standards including ISO 31000, ISO/IEC 27005, and Telecommunication Industry related Standard ITU-T X.1055 for Use case.}
    \label{fig:standard_alignment_1}
\end{figure*}

 \begin{figure*}[b]
\centering
    \includegraphics[width=0.83\linewidth]{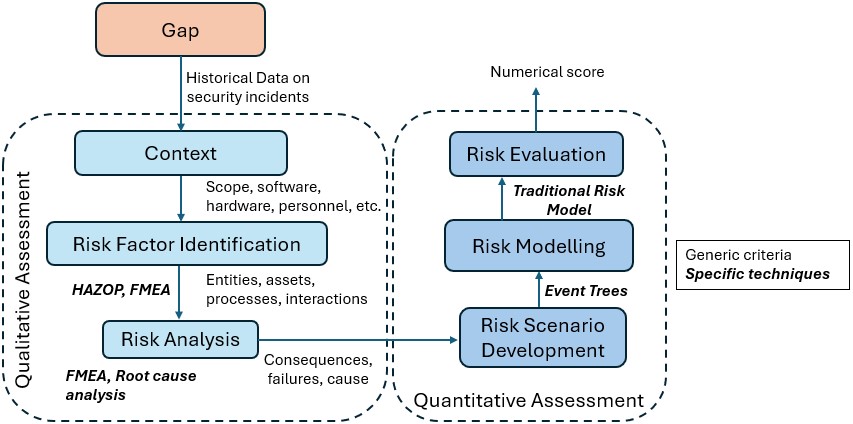}
    \caption{Proposed Security Risk Assessment Framework - TELSAFE.}
    \label{fig:generic_framework}
\end{figure*}


A few popular risk management tools in the industry include: (i) NIST Risk Management Framework (RMF), which focuses on a structured approach to identifying and managing privacy and information security risks but lacks the precision required for a quantitative risk assessment, often emphasizing qualitative risk analysis and subjective decision-making \cite{intro33, sota48}.
There are various challenges and limitations associated with the adoption and execution of RMF including extreme complexity in implementation for smaller scale, difficulty in addressing emerging threats, ensuring consistency, slow adaptation, resource limitations, etc. \cite{sota49, sota50, sota51, sota52}, (ii) Factor Analysis of Information Risk (FAIR), which while being a more quantitative framework, is typically used for assessing risks in terms of financial impact only \cite{intro34}. The limitations include time-consuming and complex implementation, data quality and timeliness problems, scalability challenges, etc. \cite{sota53}, (iii) Control Objectives for Information and Related Technology (COBIT), which provides a governance framework for managing IT systems\cite{intro35}. The challenges include resource availability for small businesses, implementation complexity, and customization etc.\cite{sota54, sota55, sota56}, and (iv) Operationally Critical Threat, Asset, and Vulnerability Evaluation (OCTAVE), which focuses on operational risk assessment, involves a more qualitative approach to evaluating assets and threats \cite{intro36}. The limitations include complex and resource-intensive integration processes, gaps in addressing all emerging threats, documentation requirements, lengthy implementation, etc. \cite{sota57, sota58, sota59}.

\section{Proposed Risk Assessment Framework}
\label{sec:proposedmethod}
\subsection{System Architecture}
\label{sec:systemarch} 

The proposed methodology aligns with the risk management frameworks outlined in ISO 31000 for generic risk management and ISO/IEC 27005 for information security-related risk management 
as illustrated in Figure \ref{fig:standard_alignment_1}. It is worth mentioning that the scope of the proposed model is limited solely to the assessment phase of the risk management process. 
It includes a qualitative assessment phase followed by a quantitative assessment phase to determine a numerical risk score. Each of these phases consists of three steps, and the description regarding 
each step has been provided below. 

\paragraph*{Historical Data} 
Organizations log security incidents in their reporting and/or analysis tools \cite{sys60}. NISTIR 8286A and 8286B provide a guide on communicating risk information utilizing risk registers and risk detail records. A risk register includes information fields regarding priority, risk description, risk category, risk response type, response cost, etc. \cite{intro22}. Depending on the focus of the risk assessment, the said data can be filtered according to time or area for more effective risk evaluations.

\subsubsection{Qualitative Assessment - TELSAFE}
\paragraph{Context} Context outlines the decision-making criteria associated with the specific organization, along with the scope of the risk management process. Depending on the security goals of an organization, the definition of the context could be completed by employing rule-based, large language model-based, or AI/ML-based methods. The context can be defined to break down the risk assessment into smaller, more manageable components for complex systems. 
        
        %
\paragraph{Risk Factor Identification} This step identifies specific entities, assets, processes, and interactions involved in a specific risk assessment/management cycle. Methods, such as interviews, HAZOP (hazard and operability), and FMEA (failure mode and effects analysis), can be employed to identify risk factors \cite{intro19}. %
An organization specifies risk factors according to its security posture and the focus of its risk management process, e.g., unpatched bugs in the software, unprotected data, lack of resource consumption limits, and weak access controls.   
\paragraph{Risk Analysis} This step covers the identification of threats due to specific gap problems and the causes of those vulnerabilities/threats, along with the corresponding impacts of those threats or vulnerabilities. Methods, such as cause-consequence analysis, scenario analysis, and root cause analysis, can be applied to analyze risk \cite{intro19}. Risk is analyzed based on the previously identified factors and the consequences they may have based on the security posture of an organization, e.g., illegitimate access to critical data due to unpatched vulnerabilities, weak authentication algorithms, weak access controls, and lack of secure data storage practices, etc. causing critical information to be exposed. 
\subsubsection{Quantitative Assessment - TELSAFE}
  
        \paragraph{Risk Scenario Development} This step utilizes the output of the previous step and applies one or a combination of techniques provided in ISO/IEC. Techniques such as event tree analysis (ETA), layers of protection analysis (LOPA), and fault tree analysis (FTA) can be utilized to develop a risk scenario for quantitative analysis \cite{intro19}. 
        Risk scenario development utilizes the vulnerabilities, their causes, and impact identified in the previous step and develops an entire risk scenario. This can be achieved by employing a number of techniques, e.g., a combination of event trees and fault trees. 
        \paragraph{Risk Modelling} The previous step can be merged with the risk modeling step when the vulnerabilities can be directly modeled according to the needs of the specific organization and selected technique. Risk modeling techniques can be employed if the developed risk scenario is not capable of capturing all desired complex interactions. Markov analysis can be implemented to model risk quantitatively \cite{intro19}. 
        Depending on the scenario developed previously, and to effectively capture intended interactions, risk modeling is performed. 
        \paragraph{Risk Evaluation} Relying on the developed risk model, the risk is evaluated depending on the requirements of an organization. This evaluation could range from a qualitative severity scale to numeric values mapped against qualitative risk bands to purely quantitative scores. Approaches, such as cost-benefit analysis (CBA), decision tree analysis, and game theory, can be implemented to model risk quantitatively \cite{intro19}. Risk is evaluated based on the focus and business demands of an organization, e.g., a risk intensity level for subjective evaluation or a numeric risk score for an objective evaluation. 
         %

\subsection{Mathematical Model}
The mathematical model focuses primarily on the 
\textit{Quantitative Assessment} as the employment of a specific approach, including event trees for this phase, has been proposed.

\begin{definition} [\textbf{Event}]
    An event encompasses all possible occurrences of interest. It can represent a specific experiment, a physical process, a natural phenomenon, a human action, a reaction to a question, and more \cite{sys62}. It is represented by \textit{e} in this manuscript. 
\end{definition}

\begin{definition} [\textbf{Outcome}]
    The outcome of an event refers to its result, which determines the consequences of the event's occurrence within the context of analysis \cite{sys62}.  It is represented by \textit{`i'} in this manuscript.
\end{definition}

\begin{definition} [\textbf{Event Outcome Space}]
    The outcome of an event can be any specific element within a defined set, where the elements are both distinct and finite. An event space has \textit{n} distinct outcomes, the outcome space encompasses all potential outcomes of an event (i.e., complete), and each individual outcome is mutually exclusive, meaning only one outcome can occur with each instance of the event \cite{sys62}. It is represented by \textit{I} in this manuscript.
\end{definition}

\begin{equation}
\label{eq:event_space}
    I = \{i_x\}, ~ ~ x = 1,2,...,n,
\end{equation}

\noindent where $I$ is the event outcome space, $i$ is an outcome, and $n$ is the total number of outcome options.

\begin{definition} [\textbf{Cartesian product of two event outcome spaces}]
    For two events $e_1$ and $e_2$ with event outcome spaces $I_1$ and $I_2$, with ${i_{1y}},~y = 1,2,...,n_1$ and ${i_{2y}},~y = 1,2,...,n_2$, respectively, the Cartesian product of the two outcome spaces is defined as the set whose elements consist of the $n_1\times n_2$ potential pairs (combinations) of the individual outcomes ${i_{1y}}$ and ${i_{2y}}$. The product of outcome spaces encompasses all potential outcomes of an event (i.e., complete), and each individual outcome is mutually exclusive, i.e., for any two unique elements of the product space $(i_{1a}i_{2b})$ and $(i_{1c}i_{2d})$, $a \neq c$ and/or $b \neq d$ \cite{sys62}. 
\end{definition}

\begin{definition} [\textbf{Cartesian product of N event outcome spaces}]
    The Cartesian product of $N$ event outcome spaces is created by taking an outcome of each space, combining it with all outcomes of every other space, and it includes all possible combinations of the outcomes of the $N$ events. It has k elements, encompasses all potential outcomes of an event (i.e., complete), and each individual outcome is mutually exclusive \cite{sys62}.
\end{definition}

\begin{equation}
\label{eq:n_outcomespace}
    I = I_1 \otimes I_2 \otimes \cdots \otimes I_N = I^{N-1}\otimes I_N,
\end{equation}

\noindent where $I^{N-1}$ represents the Cartesian product of N-1 events.

\begin{equation}
\label{eq:n_outcomespace_elements}
    k = \prod y = 1Nn_y,
\end{equation}

\noindent where $n_y$ is the number of elements on the \textit{y}th event outcome space.

\begin{definition} [\textbf{Joint event of N events}]
    A joint event of $N$ events, i.e., $e_1,e_2,...,e_N$ is the event having outcome space equivalent to the Cartesian product of N individual outcome spaces, i.e., $I_1,I_2,...,I_N$  \cite{sys62}.
\end{definition}

\begin{equation}
\label{eq:jointevent}
    e = e_1.e_2\cdots e_N.
\end{equation}

\begin{definition} [\textbf{Event base}]
    A set of $N$ events forms an event base $E$ if it includes all events whose outcomes are required to describe the state of the domain. Anything of interest should be expressible in terms of the outcomes of these $N$ events. The $N$ events that make up the event base are referred to as basic events or elements of the base \cite{sys62}. 
\end{definition}

\begin{definition} [\textbf{Outcome space of an event base}]
    The outcome space of the domain $D$ is the outcome space of the joint event formed by its event base, wherein the event outcome space of a domain is the Cartesian product of the $N$ event outcome spaces of the event base, has finite elements, encompasses all potential outcomes of an event (i.e., complete), each individual outcome is mutually exclusive, the elements are points in an N-dimensional discrete space, and a single point in the outcome space consists of $N$ individual outcomes of the $N$ events \cite{sys62}.
\end{definition}

\begin{definition} [\textbf{Event tree}]
An event tree is a visual diagram that maps out all possible outcomes of a specific event. The event's name serves as the title of the tree, and each potential outcome is represented by a branch extending from a central node \cite{sys62}.
\end{definition}

\begin{definition} [\textbf{Event tree path}]
    A path is a set of branches, one from each event in the base. Each path in an event tree represents an outcome of the joint event (or the event base). The number of paths in the tree equals the number of outcomes in the outcome space of the event base. These paths are mutually exclusive, as they correspond to mutually exclusive outcomes in the outcome space \cite{sys62}. It is represented by $\pi$ in this manuscript.
\end{definition}

\begin{equation}
\label{eq:probofoutcome}
   0\leq P[i_x] \leq 1,
\end{equation}

\noindent where $i_x$ is an outcome and $P[i_x]$ is the probability of that outcome $i_x$ and is a nonnegative real number, and 

\begin{equation}
\label{eq:sumproboutcome}
    \sum_{x=1}^{n} P[i_x]   = 1.
\end{equation}

Let an event be $e_j$,~$j=1,2,\cdots,N$ and an outcome of an event be $e_{jx}$,~$x=1,2,\cdots,n$, the probability of a path $\pi$ where events are independent can be depicted by the example of a fire in a building. The initiating event $e_1$ is the fire breaking out, the next event is related to the fire alarm activation $e_2$, followed by the activation of sprinklers $e_3$, and so on. The outcomes of the event $e_2$ include (i) $e_{21}$: Yes, where the alarm is triggered, and (ii) $e_{22}$: No, in the case the alarm is not triggered. Similarly, the outcomes related to $e_3$ will include $e_{31}$, $e_{32}$, and so on. The probability of a success path, wherein the alarm was triggered, the sprinklers were activated, the evacuation was successful, and the fire was extinguished, can be calculated by taking the product of all probabilities of individual events on this path. In general, the probability of a path $\pi$ where events are independent is computed as:
\begin{equation}
\label{eq:pathprobind}
    P[\pi] = P[e_{jx}] \times P[e_{(j+1)x}] \times P[e_{(j+2)x}] \times \cdots = \prod_{j=1}^{N} P[e_{jx}], 
\end{equation}

\noindent where $P[e_{jx}]$ is the probability of the $j$th event, $P[e_{(j+1)x}]$ is the probability of the $(j+1)$th event, and $P[e_{(j+2)x}]$ is the probability of the $(j+2)$th event \cite{addmodel1}.

Let an event be $e_j$,~$j=1,2,\cdots,N$ and an outcome of an event be $e_{jx}$, ~$j=1,2,\cdots,N$, and $x=1,2,\cdots,n$, the conditional probability of a path $P[\pi]$ is computed as:

\begin{equation}
\label{eq:pathprobcond}
    P[\pi] = P[e_{jx}] \times P[e_{(j+1)x}|e_{jx}] \times P[e_{(j+2)x}|e_{jx}\cap e_{(j+1)x}] \times \cdots
\end{equation}

\noindent where $P[e_{jx}]$ is the probability of the $j$th event, $P[e_{(j+1)x}|e_{jx}]$ is the probability of the $(j+1)$th event given the $j$th event, and $P[e_{(j+2)x}|e_{jx}\cap e_{(j+1)x}]$ is the probability of the $(j+2)$th event given the $j$th and $(j+1)$th events \cite{addmodel2}.

The impact \cite{sys63} associated with an event tree path is computed as:

\begin{equation}
    \label{eq:impact}
\gamma_\pi = 1 - [(1 - C_\gamma) \times (1 - G_\gamma) \times (1 - A_\gamma)],  
\end{equation}

\noindent where $\gamma_\pi$ is the impact or consequence associated with a path $\pi$, $C_\gamma$ is the impact on confidentiality, $G_\gamma$ is the impact on integrity, and $A_\gamma$ is the impact on availability.

The risk score $R$ associated with a path $\pi$ is computed as:
\begin{equation}
    \label{eq:risk}
R_\pi = P[\pi] \times \gamma_\pi,   
\end{equation}

\noindent where $P[\pi]$ is the likelihood, and $\gamma_\pi$ is the impact associated with a path.

    



\section{Use Case: CVE-TELSAFE}
\label{sec:use_case}

 \begin{figure}[b]
\centering
    \includegraphics[width=1.0\linewidth]{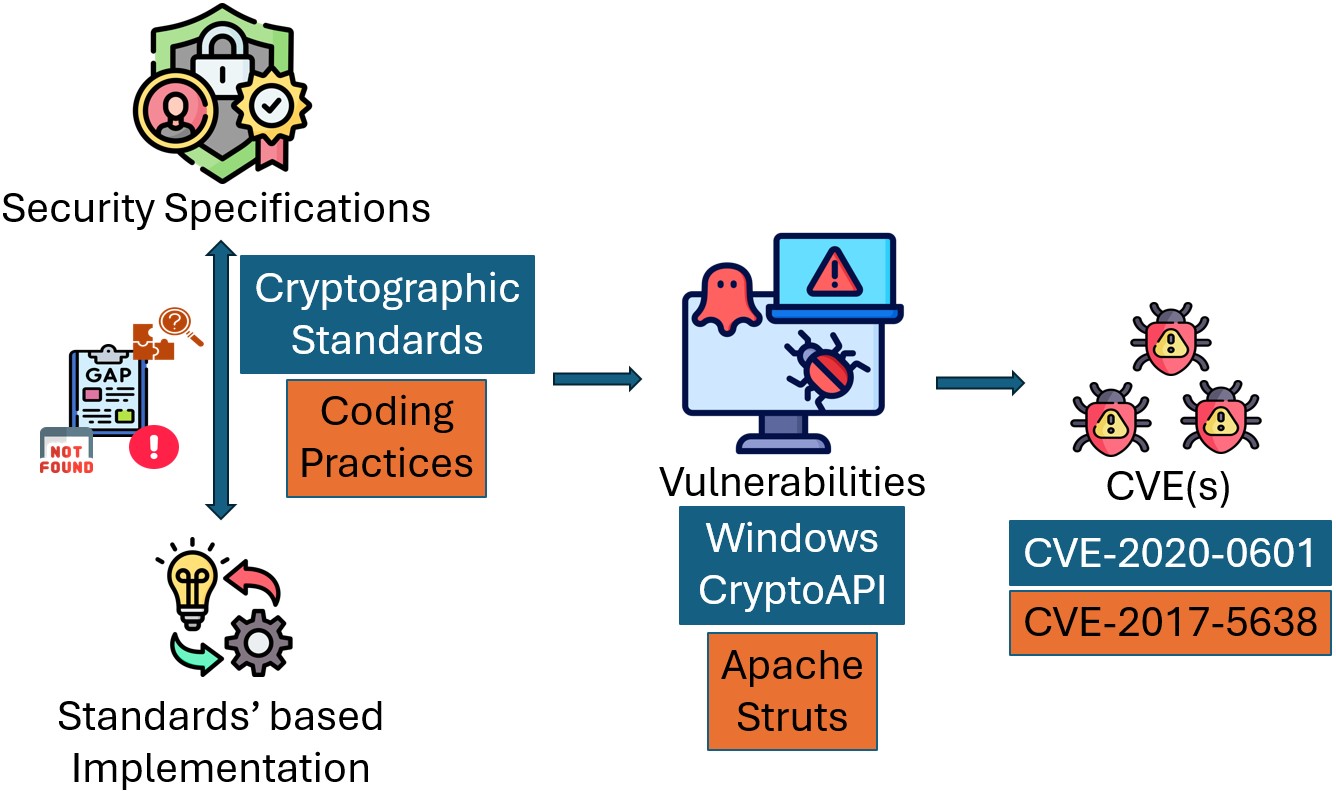}
    \caption{Standards' Non-Compliance to CVE-Classification.}
    \label{fig:gap_to_CVE}
\end{figure}

This Section covers a use case for security risk assessment 
with a primary focus on the implementation example 
of the proposed framework. Please note that the use case application 
presented here serve as a proof of concept, intended only to illustrate how the proposed framework operates on a dataset. This demonstration is particularly valuable for scenarios where the original data is proprietary or sensitive to organizations and is not available publicly.  

One application of risk assessment is in the telecommunications industry, where it can be used to evaluate the security posture of 5G/6G networks. Among other standardization organizations mentioned earlier, the International Telecommunication Union (ITU) is actively engaged in improving the security in the telecommunications sector \cite{intro1, intro2}. With regard to risk assessment in the telecommunications industry, there is no evidence of a sector-based risk assessment framework being successfully employed. There are two such ongoing projects: (i) TISRIM, which was proposed as a national security risk management framework to ensure compliance with national and European regulations for TSPs. This TISRIM framework consists of two components: a security risk management tool, which can be employed by TSPs, and an analysis tool for the regulatory authority to collect and evaluate the risk management reports submitted by TSPs. Following the first regulatory cycle, limitations of the proposed framework have been identified by the National Regulatory Authority of Luxembourg and the regulated entities. The activity plan to address these limitations has also been presented; however, the current implementation status of the envisaged framework is unclear \cite{intro37}, and (ii) Telecommunications Risk and Resilience Profile Pestle and Gap Analysis, an ongoing project that aims to create a comprehensive risk and resilience profile for the telecommunications sector using an all-hazards approach and to establish an independent evidence base to inform future policy decisions \cite{intro38}. The proposed methodology also aligns with the risk management frameworks outlined in  ITU-T X.1055 for telecommunications industry-related risk management, in addition to ISO 31000 for generic risk management and ISO/IEC 27005 for information security-related risk management, as illustrated in Figure \ref{fig:standard_alignment_1}. 

 \begin{figure*}[t]
\centering
    \includegraphics[width=0.78\linewidth]{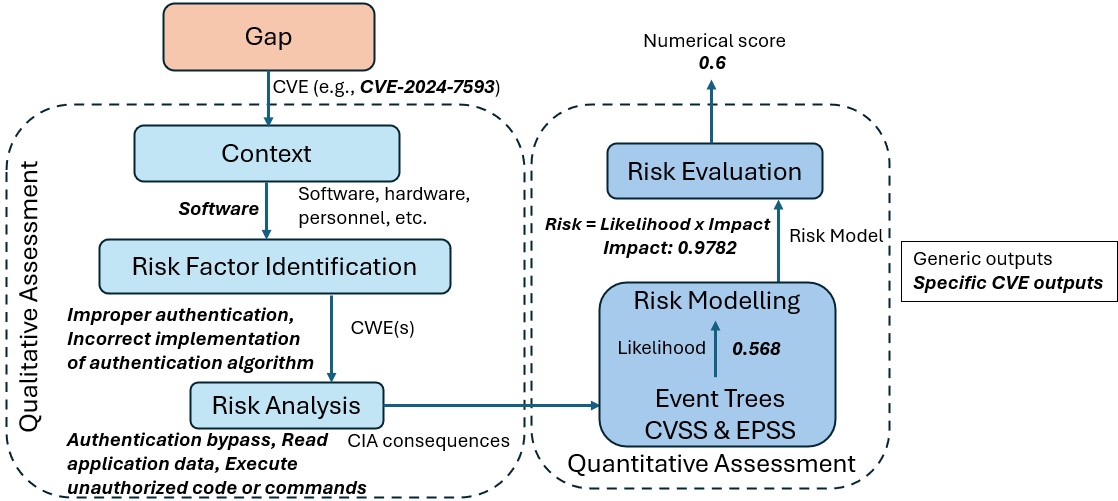}
    \caption{CVE-based Use Case Security Risk Assessment Framework (CVE-TELSAFE).}
    \label{fig:specific_framework}
\end{figure*}

As mentioned before, the gaps 
between established security standards and their practical implementation may create vulnerabilities in systems. These vulnerabilities are subsequently identified, documented, and reported, provided that they are significant enough to the Common Vulnerabilities and Exposures (CVE) program for awareness and remediation \cite{usecase64} as depicted in Figure \ref{fig:gap_to_CVE}. With CVE information, the owner of the asset gains a better understanding of the security risks \cite{usecase65}. The National Vulnerability Database (NVD) is one of the most extensive repositories of known vulnerabilities. It catalogs software and hardware vulnerabilities using the Common Platform Enumeration (CPE) format. These vulnerabilities are recorded in the Common Vulnerabilities and Exposures (CVE) format and are assessed using the Common Vulnerability Scoring System (CVSS), which rates them on a scale of ``None, Low, Medium, High, and Critical''~\cite{usecase66, usecase67, usecase68}. Another scoring system known as the Exploit Prediction Scoring System (EPSS) is a data-driven framework designed to estimate the probability that a software vulnerability will be exploited in real-world scenarios. It employs up-to-date information about the threat from CVE and real-world exploit data. The EPSS model generates a probability score ranging from 0 to 1, with higher scores indicating a higher likelihood that a vulnerability is exploited \cite{usecase69}. Together, these two scoring mechanisms, i.e., CVSS and EPSS, can be employed for a more informed evaluation of risk associated with individual CVEs \cite{usecase70, usecase71}. As mentioned by NIST \cite{usecase72}, the CVSS score itself is not a measure of risk. However, different metrics/parameters involved in CVSS can help estimate the likelihood of an event and the corresponding impact \cite{intro29}. 
Table \ref{tab:cvesrelevance} presents some examples of the direct/indirect relevance of CVE(s) to (i) the gaps between established security standards and their practical implementation, or/and (ii) the telecommunications industry.
Figure \ref{fig:specific_framework} illustrates the step-wise flow diagram of the CVE-based security risk assessment.

\begin{table}
\begin{center}

\caption{Relevance of CVE(s) 
- Examples. }%
\label{tab:cvesrelevance}
{\renewcommand{\arraystretch}{0.05}
\tiny{
\begin{tabular}{p{1.3cm}p{0.6cm}p{2.4cm}p{2.3cm}}
\toprule
\centering{\textbf{CVE ID}} & \centering{\textbf{(i), (ii)\tablefootnote{CVE relevance to (i) the gaps between established security standards and their practical implementation, and (ii) the telecommunications industry.}}} & \textbf{Description} & \textbf{Gap} \\
\tabularnewline
\midrule
\centering{CVE-2024-3845} & \centering{(i), (ii)} & Google Chrome mixed content policy bypass 
& Inconsistency between implementation and documented design 
\\\tabularnewline[0.2cm]

\centering{CVE-2023-30590} & \centering{(i), (ii)} & Diffie-Hellman public and private key documentation mismatch & Discrepancy between the documented behavior and the actual implementation \\\tabularnewline[0.2cm]

\centering{CVE-2022-4304} & \centering{(i), (ii)} & OpenSSL RSA decryption timing side-channel
& Implementation vulnerability, observable discrepancy \\\tabularnewline[0.2cm]

\centering{CVE-2023-32342} & \centering{(i), (ii)} & IBM GSKit RSA decryption timing flaw
& Implementation vulnerability, observable discrepancy  \\\tabularnewline[0.2cm]

\centering{CVE-2022-3703} & \centering{(i), (ii)} & Etic Telecom remote access server 4.5.0 firmware verification bypass 
& Implementation vulnerability 
\\\tabularnewline[0.2cm]

\centering{CVE-2023-47610} & \centering{(i), (ii)} & Telit Cinterion EHS5/6/8 remote code execution & Implementation vulnerability 
\\\tabularnewline[0.2cm]

\centering{CVE-2023-20198} & \centering{(i), (ii)} & Cisco IOS XE web UI privilege escalation & Discrepancy between best practices defined in security standards and their practical implementation \\\tabularnewline[0.4cm]



\centering{CVE-2025-2185} & \centering{(i), (ii)} & ALBEDO Telecom Net.Time - PTP/NTP clock vulnerability
&  Implementation vulnerability
\\\tabularnewline[0.2cm] 



\centering{CVE-2024-4225} & \centering{(i), (ii)} & Security vulnerabilities in NetGuardian DIN RTU web interface by DPS Telecom
&  Implementation vulnerability
\\\tabularnewline[0.2cm]

\centering{CVE-2018-0171} & \centering{(i), (ii)} &  Cisco IOS software and Cisco IOS XE software smart install vulnerability 
&  Improper implementation of security principles 
\tabularnewline




\bottomrule
\end{tabular}
}}
\end{center}
\end{table}


\begin{table*}[b]
\begin{center}

\caption{Structure of the Dataset $\mathcal{D}$ after Pre-processing.}%
\label{tab:moddatastructure}
{\renewcommand{\arraystretch}{0.05}
\tiny{
\begin{tabular}{p{1.2cm}p{0.4cm}p{0.4cm}p{0.8cm}p{0.5cm}p{0.4cm}p{0.6cm}p{0.4cm}p{0.5cm}p{0.8cm}p{0.6cm}p{0.8cm}p{0.4cm}p{0.8cm}p{0.6cm}p{0.6cm}p{0.7cm}}
\toprule
\textbf{\makecell{\\CVE ID}} & \textbf{\makecell{Base\\Severity}} & \textbf{\makecell{Base\\Score}} & \textbf{\makecell{Exploitability\\Score}} & \textbf{\makecell{Impact\\Score}} & \textbf{\makecell{EPSS\\Score}} & \textbf{\makecell{EPSS\\Percentile}}  & \textbf{\makecell{CISA\\Kev}} & \textbf{\makecell{Attack\\Vector}} 
& \textbf{\makecell{Attack\\Complexity}} & \textbf{\makecell{Privileges\\Required}} & \textbf{\makecell{User\\Interaction}} & \textbf{\makecell{Scope}} & \textbf{\makecell{Confidentiality\\Impact}} & \textbf{\makecell{Integrity\\Impact}} & \textbf{\makecell{Availability\\Impact}} & \textbf{\makecell{Published\\Date}}\\
\tabularnewline
\midrule
\centering{CVE-1999-0199} & \centering{3} & \centering{9.8} & \centering{3.9} & \centering{5.9} & \centering{0.00729} & \centering{0.81305} & \centering{0} & \centering{3} & \centering{0} & \centering{0} & \centering{0} & \centering{0} & \centering{2} & \centering{2} & \centering{2} & \centering{2020-10-06T13:15Z}\\\tabularnewline

\centering{CVE-1999-0236} & \centering{2} & \centering{7.5} & \centering{3.9} & \centering{3.6} & \centering{0.0028} & \centering{0.69174} & \centering{0} & \centering{3} & \centering{0} & \centering{0} & \centering{0} & \centering{0} & \centering{2} & \centering{0} & \centering{0} & \centering{1997-01-01T05:00Z}\\\tabularnewline[-0.2cm]

\centering{~\vdots} & \centering{} & \centering{} & \centering{} & \centering{} & \centering{} & \centering{} & \centering{} & \centering{} & \centering{} & \centering{} & \centering{} & \centering{} & \centering{} & \centering{} & \centering{} & \centering{-}\\\tabularnewline[0.3cm]

\centering{CVE-2020-8578} & \centering{0} & \centering{3.3} & \centering{1.8} & \centering{1.4} & \centering{0.00044} & \centering{0.14123} & \centering{0} & \centering{1} & \centering{0} & \centering{1} & \centering{0} & \centering{0} & \centering{1} & \centering{0} & \centering{0} & \centering{2021-02-08T22:15Z}\\\tabularnewline

\centering{CVE-2020-8579} & \centering{2} & \centering{7.5} & \centering{3.9} & \centering{3.6} & \centering{0.00103} & \centering{0.43754} & \centering{0} & \centering{3} & \centering{0} & \centering{0} & \centering{0} & \centering{0} & \centering{0} & \centering{0} & \centering{2} & \centering{2020-10-27T14:15Z}\\\tabularnewline[-0.2cm]

\centering{~\vdots} & \centering{} & \centering{} & \centering{} & \centering{} & \centering{} & \centering{} & \centering{} & \centering{} & \centering{} & \centering{} & \centering{} & \centering{} & \centering{} & \centering{} & \centering{} & \centering{-}\\\tabularnewline[0.3cm]

\centering{CVE-2024-9996} & \centering{2} & \centering{7.8} & \centering{1.8} & \centering{5.9} & \centering{0.00066} & \centering{0.31062} & \centering{0} & \centering{1} & \centering{0} & \centering{0} & \centering{1} & \centering{0} & \centering{2} & \centering{2} & \centering{2} & \centering{2024-10-29T22:15Z}\\\tabularnewline

\centering{CVE-2024-9997} & \centering{2} & \centering{7.8} & \centering{1.8} & \centering{5.9} & \centering{0.00066} & \centering{0.31062} & \centering{0} & \centering{1} & \centering{0} & \centering{0} & \centering{1} & \centering{0} & \centering{2} & \centering{2} & \centering{2} & \centering{2024-10-29T22:15Z}\tabularnewline
\bottomrule
\end{tabular}
}}
\end{center}
\end{table*}

\subsection{Dataset}
\label{sec:dataset}
The data maintained by an organization (e.g., telecommunications operator) related to the logged security incidents or recorded risk registers is not available publicly \cite{sys60}. Accordingly, the dataset employed for this use case is taken from \emph{kaggle}\footnote{https://www.kaggle.com/datasets/francescomanzoni/vulnerability-management-datasets}, 
which consolidates data from (i) NVD, (ii) EPSS, and (iii) CISA-KEV(Cybersecurity and Infrastructure Security Agency-Known Exploited Vulnerabilities). The said dataset contains 155,578 samples of vulnerability data (i.e., 154,528 samples for vulnerabilities that were not exploited and 1,050 samples for exploited vulnerabilities). 
Please note that this data has not been acquired by the source as data for gaps between standards and implementation, however, it can be linked to the gaps as the reported CVEs include vulnerabilities reported due to these gaps, e.g., CVE-2017-5638 and CVE-2020-0601 included in the said CVE dataset were created after vulnerabilities due to gaps in coding practices and cryptographic standards
respectively were discovered (refer to Figure \ref{fig:gap_to_CVE}).



\subsubsection{Original Dataset}
The original dataset contains (i) base severity levels, i.e., low, medium, high, and critical; (ii) CISA KEV information in terms of true, and false; (iii) attack vector types as physical, local, adjacent-network, and network, (iv) attack complexity as low, and high, (v) privileges required as none, low and high, (vi) user interaction as none, and required, (vii) scope in terms of change, and unchanged, and (viii) CIA impact is categorized as none, low, and high for each component.

\subsubsection{Pre-processed Dataset ($\mathcal{D}$)}
To facilitate computations, the original dataset has been pre-processed to assign numerical values to each parameter, and the modified dataset $\mathcal{D}$ includes (i) base severity levels as 0 for low, 1 for medium, 2 for high, and 3 for critical, (ii) CISA KEV information as 1 for true, and 0 for false, (iii) attack vector types as 0 for physical, 1 for local, 2 for adjacent-network, and 3 for network, (iv) attack complexity as 0 for low, and 1 for high, (v) privileges required as 0 for none, 1 for low and 2 for high, (vi) user interaction as 0 for none, and 1 for required, (vii) scope in terms of 0 for unchanged, and 1 for changed, and (viii) CIA impact is categorized as 0 for none, 1 for low, and 2 for high for each component. Table \ref{tab:moddatastructure} illustrates the structure and a few samples of the pre-processed dataset.

\subsection{Qualitative Assessment CVE-TELSAFE}
The qualitative assessment part has been carried out manually by utilizing the information available at \cite{NVD_2024_7593}. 

\subsubsection{Context}
The scope of risk management includes software, hardware, personnel, etc. Figure \ref{fig:specific_framework} shows an example output of context for CVE-2024-7593. %
 %
\subsubsection{Risk Factor Identification}
Risk factors are specified according to the attributes included in the CVSS model, e.g., requirement of privileges, access type, and attack complexity. Figure \ref{fig:specific_framework} presents an example output of this step for CVE-2024-7593. 
\subsubsection{Risk Analysis} 
Risk is analyzed relying on the type of vulnerability due to the gap, risk factors, and the influence on confidentiality, integrity, and availability. Figure \ref{fig:specific_framework} exhibits an example output of this step for CVE-2024-7593. %

\subsection{Quantitative Assessment CVE-TELSAFE}
   
\subsubsection{Risk Scenario Development} 
This probabilistic modeling step employs the concept of event trees on the probability influencing parameters of a CVE to compute the numeric value of the likelihood factor. Algorithm \ref{algorithm:RiskAssess} delineates the development of event trees for this particular use case. Each likelihood influencing parameter (eight in total) in $\mathcal{D}$  has been considered as an event to form an event tree for likelihood computations $P[\pi]$. Occurrences of each unique outcome of each event have been calculated to compute the probability of each outcome $P[i_x]$. The following parameters have been considered as likelihood influencing attributes: base score, exploitability score, EPSS percentile, attack vector, attack complexity, privileges required, user interaction, and scope. 
\subsubsection{Risk Modeling} 
In this case, risk modeling and probabilistic modeling steps can be combined to compute numerical values for the two constituents of risk, i.e., likelihood $P[\pi]$ and impact $\gamma^{\textup{N}}_{\pi}$. The impact level has been computed by incorporating impacts on confidentiality, integrity, and availability, whereas the impact score $\gamma^{\textup{N}}_{\pi}$ has been derived from the CVSS impact score. 
\subsubsection{Risk Evaluation}
Numeric risk score $R_\pi$ along with a normalized numeric risk value $R^{\textup{N}}_{\pi}$ is computed by taking into consideration the previously evaluated likelihood $P[\pi]$ and impact factors $\gamma^{\textup{N}}_{\pi}$. 
%

\subsubsection{Outputs}
\label{sec:results_analysis}
The step-wise outputs of (i) the proposed framework CVE-TELSAFE (which utilizes a hybrid approach) and (ii) a qualitative approach (performed manually) based on ISO 31000, for the same CVE (i.e., CVE-2024-7593) have been compared in Table \ref{tab:compare_approaches}. 


\begin{table*}[t]
\centering
\caption{Comparison Risk Assessment 
- Qualitative Approach vs Proposed Hybrid Approach (CVE-2024-7593)}%
\label{tab:compare_approaches}
{\renewcommand{\arraystretch}{0.5}
\scriptsize{
\begin{tabular}{p{3cm}p{4.5cm}p{7.5cm}}
\toprule
\textbf{Step} & \textbf{Qualitative Output} & \textbf{Proposed Hybrid Output}\\ 
\tabularnewline
\midrule
Context & -- & Software \\\tabularnewline 

Risk Identification & Risk factors
& Risk factors related to vulnerabilities (Improper authentication, Incorrect implementation of authentication algorithm)
 \\\tabularnewline 

Risk Analysis & Consequences and Likelihood 
 & Consequences in terms of confidentiality, integrity, and availability (Authentication bypass, Read application data, Execute unauthorized code or commands)
  \\\tabularnewline 

Risk Scenario Development & -- &  Event trees $\rightarrow$ Likelihood (0.568) 
\\\tabularnewline 

Risk Modeling & -- & Likelihood (0.568), Impact (0.9782) $\rightarrow$ Risk 
 \\\tabularnewline 

Risk Evaluation & Risk level only
&  Risk score (numeric) (0.6)$\rightarrow$ Risk level (Risky)
\\\tabularnewline \bottomrule
\end{tabular}
}}
\end{table*}

\begin{algorithm}[!tb]
\caption{Risk Assessment: CVE-TELSAFE}
\label{algorithm:RiskAssess}
\SetInd{0.25em}{0.1em} 
\footnotesize{
\textbf{Input}: Dataset $\mathcal{D}$. \\
\textbf{Output}: Numeric Risk Scores $R_\pi$. \\
\For{$t \gets cycle$}{
    Initialize timer $t_i$\;
    $dataArray \gets \mathcal{D}$\;
    $LhdCol \gets \textup{cell}(\text{size}(dataArray,1), \text{size}(N,2))$\;
    $q \gets 1$\;
    \For{$colx \in [1 \dots N]$}{ 
        $LhdCol(:,q) \gets dataArray(:, colx)$\;
        $q \gets q+1$\;
    }
    \For{$j \gets 1 \text{ to } N$}{
        $colData \gets LhdCol(:, j)$\;
        $(unqVal, idx) \gets \text{unique}(colData)$\;
        \For{$x \gets 1 \text{ to } \textup{numel}(\text{unqVal})$}{
            $value \gets unqVal(x)$\;
            $occurrences \gets \text{sum}(\text{idx} == x)$\;
            $P[i_x] \gets \frac{occurrences}{\text{numel}(colData)}$\;
            $P[e_{jx}](\text{idx} == x, j) = P[i_x]$\;
        }
    }
    \For{$row \gets 1 \text{ to } \textup{size}(dataArray, 1)$}{
        $P[\pi] \gets \prod_{j=1}^{N} P[e_{jx}]$\;
        $\gamma^{\textup{N}}_{\pi} \gets \frac{\gamma_\pi - \gamma_{\min_T}}{\gamma_{\max_T} - \gamma_{\min_T}}$\;
        $R_\pi \gets P[\pi] \times \gamma_\pi$\;
        $R^{\textup{N}}_{\pi} \gets \frac{R_\pi - R_{\min_T}}{R_{\max_T} - R_{\min_T}}$\;
    }
    End timer $t_f$\;
}
}
\end{algorithm}

\begin{table*}[t]
\centering
\caption{Computation based Comparison With Existing Literature}%
\label{tab:compare_literature_results}
{\renewcommand{\arraystretch}{0.5}
\scriptsize{
\begin{tabular}{p{2.0cm}p{14.5cm}}
\toprule
\textbf{Framework} & \textbf{Comparison} \\
\tabularnewline
\midrule
CyRiPred \cite{intro31} & (i) utilizes CVE based dataset, (ii) computed values have been provided but not against CVEs like ours but year wise, (iii) risk scores are computed based on categories, e.g., transport layer security, cross-site scripting, SQL injection, etc., (iv) the goal is to categorize into topics on Wikipedia, (v) not related to telecommunication industry
\\\tabularnewline
\cite{intro29}  & (i) utilizes CVSS-based dataset, (ii) no dataset has been made available, (iii)  our dataset does not have values against all the parameters used in this model, (iv) no computed values have been provided by the authors 
for comparison, (v) not related to the telecommunication industry
\\\tabularnewline
\cite{sota39} & (i) utilizes CWE based (10 only) dataset, (ii) comparison is not possible as the computed values have been provided but not against CVEs like ours,(iii) risks of specific attack categories (06), e.g., improper authorization, cross-site scripting, SQL injection, etc., (iv) not related to telecommunication industry
\\\tabularnewline
\cite{intro30}  & (i) not a CVSS/CVE based approach, (ii) no dataset has been made available, (iii) the parameters used are not available in our dataset, (iv) no computed values have been provided by the authors 
for comparison, (v) not related to telecommunication industry  \\\tabularnewline
AFPr-AM \cite{sota40} & (i) via questionnaires, (ii) the parameters of the dataset do not match, (iii) no computed values have been provided by the authors 
for comparison, (iv) not related to the telecommunication industry
\\\tabularnewline
\cite{intro32} & (i) studies impacts of AI on risk management via questionnaires, (ii) no computed values have been provided by the authors for comparison, (iii) not related to the telecommunication industry  \\\tabularnewline
\bottomrule
\end{tabular}
}}
\end{table*}

\begin{table*}[t]
\centering
\caption{Framework \& Computation based Comparison With Existing Industry Tools}%
\label{tab:compare_tools_results}
{\renewcommand{\arraystretch}{0.5}
\scriptsize{
\begin{tabular}{p{2.0cm}p{13.7cm}}
\toprule
\textbf{Framework} & \textbf{Comparison} \\
\tabularnewline
\midrule
RMF \cite{intro33} & (i) no quantitative results to be compared, (ii) no specific techniques but guidelines, (iii) subjective
\\\tabularnewline 
FAIR \cite{intro34} & (i) quantitative, (ii) however, the focus is on financial risk assessment only
\\\tabularnewline
COBIT \cite{intro35} & (i) focuses only on business management risk and information governance, (ii) no quantitative results to be compared, (iii) subjective \\\tabularnewline
OCTAVE \cite{intro36} &  (i) focuses only on business operational risk, (ii) no quantitative results to be compared, (iii) subjective
\\\tabularnewline
EU \cite{intro37} & (i) on-going, (ii) no results to be compared 
\\\tabularnewline
ANU \cite{intro38} & (i) on-going, (ii) no results to be compared 
\\\tabularnewline
\bottomrule
\end{tabular}
}}
\end{table*}

\subsubsection{Discussion}
\label{sec:discussion}

\paragraph{Contributions} As mentioned earlier, the proposed TELSAFE framework aimed to (i) quantify risk, (ii) eliminate subjectivity, and (iii) yield consistent results. 
Quantification: The proposed framework yielded quantified and numerical risk scores by applying equations~\ref{eq:probofoutcome}--\ref{eq:risk}. Subjectivity: The proposed framework did not require expert opinions to assign any membership functions. Consistency: The proposed framework will yield the same results regardless of the user, as the attributes are data-driven. 

\paragraph{Evaluation}
The evaluated risk scores by employing Algorithm \ref{algorithm:RiskAssess} against the CVEs included in the CVE dataset (balanced/unbalanced) can be made available on request.

\paragraph{Validation}
The most significant challenge when devising a risk assessment framework is the unavailability of a labeled dataset, i.e., a dataset containing ground truth values for (i) risk assessment, (ii) likelihood, and (iii) impact. 
Moreover, the security incident log or risk register data of organizations is not publicly available \cite{sys60}. It hinders the feasibility of risk score validation through traditional ways, i.e., by comparing the results with the established ground truth. Therefore, conventional validation methods relying on labels or ground truth assessments may not be applicable.

\paragraph{Comparison}

Table \ref{tab:compare_approaches} provides a stepwise comparison of the proposed hybrid risk assessment approach (TELSAFE) to the qualitative approach. The qualitative approach has 3 steps involved in the risk assessment based on ISO 31000, whereas the proposed risk assessment approach consists of a total of 6 steps. The first step \textit{context} is unique to the proposed approach, which implies that by employing the traditional qualitative method, there is no option to segment or divide the entire risk assessment into smaller, manageable parts. The next step, i.e., \textit{risk identification} is the same in both approaches as the proposed framework also applies qualitative assessment method for this step. While assessing risk by using a qualitative technique alone, the \textit{risk analysis} step comprises understanding both the consequences and the likelihood. Moreover, it is a common practice in these techniques to estimate likelihood and consequences in terms of severity or a scale (e.g., low, medium, high). In the proposed framework, the \textit{risk analysis} process has been further divided into \textit{risk analysis}, \textit{risk scenario development}, and \textit{risk modeling}. This division into smaller and simplified steps helps in performing the entire process in a more manageable way. The \textit{risk analysis} step relies on the qualitative method and assesses the risks in terms of CIA (confidentiality, integrity, and availability) based consequences. The subsequent steps in the proposed hybrid framework lean on quantitative assessment techniques. The \textit{risk scenario development} step creates an event tree to measure the likelihood, whereas the \textit{risk modeling} step formulates the risk considering the parameters of likelihood and impact. The last step in both approaches evaluates the risk; however, the qualitative approach only evaluates the risk in terms of severity or a risk level, whereas the proposed framework estimates the risk in both the numerical value and a risk level.

Table \ref{tab:compare_literature_results} presents the reasons why result comparisons with each existing work are not feasible. Comparisons can only be made with existing works that compute risk scores quantitatively, provide risk calculations to be compared with, employing CVSS metrics similar to the proposed framework, and provide the dataset utilized. Table \ref{tab:compare_tools_results} outlines the reasons why comparing results with each existing tool is not achievable. Comparisons can not be made with tools that only include guidelines and not specific techniques to be able to compute risk values for comparison, utilize qualitative evaluation, and are subjective.

\paragraph{Core Findings}
Understanding an organization's requirements is essential when developing a tailored risk assessment model for it. Different techniques can be applied in each phase of the proposed risk assessment process, depending on the requirements of an organization. Moreover, determining the scope of vulnerabilities is vital for modeling risk. Furthermore, defining the acceptable risk is challenging and crucial for decision making. 

\section{Conclusion \& Future Work}
\label{sec:conclusion}
This paper has presented a security risk assessment framework for gaps between the standards' specifications and their implementation. These gaps may lead to chain events/failures, making a structured approach essential for assessing the probability of such event sequences. Assessing these gaps and prioritizing risks associated with them based on likelihood is of significant importance, especially for enterprise networks, complex systems, and critical infrastructure such as the telecommunications industry. Moreover, incorporating specific techniques instead of general guidelines is also crucial when devising a risk management model. 
By employing probabilistic modeling, we have evaluated quantified risk scores associated with security vulnerabilities. A use case that demonstrates the application of the proposed risk assessment framework 
on a CVE dataset has also been included. The CVE dataset is utilized due to the lack of publicly available datasets related to the real security incident logs or risk registers maintained by an organization (such as a telecommunications provider). The CVE dataset includes a range of vulnerabilities that include examples of those that arise from gaps between standards and implementation.

In the future, the authors intend to carry out validation by employing 
model-checking approaches. Moreover, we intend to present a more effective risk assessment procedure by combining the strengths of multiple techniques, e.g., probabilistic modeling and AI-based approaches, while minimizing their attack surface-related weaknesses.


\section{Acknowledgments}
This research paper is conducted under the 6G Security Research and Development Project, as led by the Commonwealth Scientific and Industrial Research Organization (CSIRO), through funding appropriated by the Australian Government’s Department of Home Affairs. This paper does not reflect any Australian Government policy position. For more information regarding this Project, please refer to \url{https://research.csiro.au/6gsecurity/}.

\balance

\end{document}